\documentclass{aa}  

\usepackage{graphicx}
\usepackage{txfonts}
\usepackage{url}
\usepackage{enumerate}
\usepackage{color}
\usepackage{hyperref}
\usepackage{amsmath}
\usepackage{verbatim}
\usepackage{footmisc}
\usepackage[hyphenbreaks]{breakurl}
\makeatletter \g@addto@macro\UrlSpecials{\do\!{\newline}}

\usepackage{natbib}
\bibpunct{(}{)}{;}{a}{}{,} 

\newcommand{\spotter}{{\sc neoranger}}

\begin{document} 

\newcommand\numberthis{\addtocounter{equation}{1}\tag{\theequation}}

\title{Monitoring near-Earth-object discoveries for imminent impactors}

\author{Otto Solin\inst{\ref{uh}} \and Mikael Granvik\inst{\ref{uh},\ref{ltu}}}

\institute{Department of Physics, P.O. Box 64, FI-00014 University of
  Helsinki, Finland, \email{otto.solin@helsinki.fi}\label{uh} \and
  Department of Computer Science, Electrical and Space Engineering, Lule\aa{} University of Technology, Box 848, S-98128 Kiruna, Sweden\label{ltu}}

\date{}


\abstract
{}
{We present an automated system called \protect \spotter{} that
  regularly computes asteroid-Earth impact probabilities for
  objects on the Minor Planet Center's (MPC) Near-Earth-Object
  Confirmation Page (NEOCP) and sends out alerts of imminent impactors
  to registered users. In addition to potential Earth-impacting
  objects, \protect \spotter{} also monitors for other types of
  interesting objects such as Earth's natural temporarily-captured
  satellites.}
{The system monitors the NEOCP for objects with new data and solves,
  for each object, the orbital inverse problem, which results in a
  sample of orbits that describes the, typically highly-nonlinear,
  orbital-element probability density function (PDF). The
  PDF is propagated forward in time for seven days and the impact
  probability is computed as the weighted fraction of the sample
  orbits that impact the Earth.}
{The system correctly predicts the then-imminent impacts of
  2008~TC$_3$ and 2014~AA based on the first data sets
  available. Using the same code and configuration we find that the
  impact probabilities for objects typically on the NEOCP, based on eight
  weeks of continuous operations, are always less than one in ten million,
  whereas simulated and real Earth-impacting asteroids always have an
  impact probability greater than 10\% based on the first two
  tracklets available.}
{}



\authorrunning{O.~Solin and M.~Granvik}
\maketitle

%

\section{Introduction}\label{sec:intro}

We have set up an automated system called \protect \spotter{} that is
built on OpenOrb\footnote{\url{https://github.com/oorb/oorb}.}
\citep{2009M&PS...44.1853G} and computes asteroid-Earth impact
probabilities for new or updated objects on the Near-Earth-Object
Confirmation Page
(NEOCP\footnote{\url{https://www.minorplanetcenter.net/iau/NEO/toconfirm\_tabular.html}.})
provided by the Minor Planet Center (MPC). The orbit-computation
methods used by \protect \spotter{} are optimized for cases where the
amount of astrometry is scarce or it spans a relatively short
time span.

The main societal benefit of monitoring for imminent impactors is that
it allows for a warning to be sent out in case of an immiment impact
that can no longer be prevented. Two well-known examples of
damage-causing impacts are the Tunguska \citep{1993Natur.361...40C}
and Chelyabinsk \citep{2013Sci...342.1069P} airbursts. Neither of
these two asteroids hit the surface of the Earth intact but instead
disrupted in the atmosphere with the resulting shock waves being
responsible for the damage on the ground. While the Tunguska event in
1908 destroyed a wide forested area, it did not cause any injuries
because it happened in a sparsely populated area. The Chelyabinsk
event in 2013, on the other hand, took place in an urban area injuring
many residents and causing damage to buildings. The Chelyabinsk
asteroid was undetected before its entry into the atmosphere, because
it arrived from the general direction of the Sun.

The primary scientific benefit of discovering asteroids prior to their
impact with the Earth is that it allows for the characterization of
the impactor in space prior to the impact event. Cross-correlating
spectroscopic information and the detailed mineralogical information
obtained from the meteorites found at the impact location allows us to
extend the detailed mineralogical information to all asteroids by
using the spectroscopic classification of asteroids as a proxy. Only
two asteroids have been discovered prior to entry into the Earth's
atmosphere, 2008~TC$_3$ \citep{2009Natur.458..485J} and 2014~AA
\citep{2016Icar..274..327F}. Both asteroids were small (no more than
five meters across) and hence discovered only about 20 hours before
impact. Whereas hundreds of observations were obtained of 2008~TC$_3$,
only seven observations were obtained of 2014~AA. The impact location
and time were accurately predicted in advance for 2008~TC$_3$ whereas
for 2014~AA the accurate impact trajectory was reconstructed
afterwards using both astrometric observations and infrasound
data. Meteorite fragments were recovered across the impact location
for 2008~TC$_3$ whereas no meteorite fragments could be collected for
2014~AA, which fell into the Atlantic Ocean.  Spectroscopic
observations in visible wavelengths were obtained for 2008~TC$_3$
suggesting that the object is a rare F-type. Mineralogical analysis of
one of the fifteen recovered meteorites showed it to be an anomalous
polymict ureilite and thereby established a link between F-type
asteroids and ureilites \citep{2009Natur.458..485J}.

Let us make the simplifying assumption that meteorites will be
produced in the impacts of all asteroids larger than one meter in
diameter.  \citet{2002Natur.420..294B,2013Natur.503..238B} estimate
that five-meter-diameter or larger asteroids (that is, similar to
2008~TC$_3$) impact the Earth once every year.  Since about 70\% of
the impacts happen above oceans and roughly 50\% of the impactors will
come from the general direction of the Sun, our crude estimate is that
an event similar to the impact of 2008~TC$_3$ (pre-impact
spectroscopic observations and discovery of meteorites) will happen
approximately twice in a decade. Since not all of the incoming
asteroids are detected, our estimate is to be considered
optimistic. This is also supported by the fact that there has not been
another 2008~TC$_3$ -like event in the past decade. These rates imply
that it will take at least a millennium to get a statistically
meaningful sample of, say, ten events for each of the approximately 20
Bus-DeMeo spectral classes. However, the rate of events can be
increased if the surveys could detect smaller objects prior to their
impact.

The Asteroid Terrestrial-impact Last Alert System
\citep[ATLAS,][]{2018arXiv180200879T} seeks to provide a warning time
in the order of tens of hours for a Chelyabinsk-size object. ATLAS
currently consists of two telescopes in the Hawaiian Islands with
plans to build additional observatories in other parts of the
world. The cameras of these two telescopes have extremely wide fields
of view compared to other contemporary Near-Earth-Object (NEO) survey systems, and they
scan the sky in less depth but more quickly producing up to 25,000
asteroid detections per night. The European Space Agency's (ESA) fly-eye telescope is a similar
concept with very large field of view. First light is expected in 2018
and it will be the first telescope in a proposed future European
network that would scan the sky for NEOs. We stress that both ATLAS 
and ESA's fly-eye telescopes have been funded primarily to reduce the 
societal risks associated to small asteroids impacting the Earth.

Pushing the detection threshold down to one-meter-diameter asteroids
will increase the discovery rate for Earth impactors by a factor of
40. A factor of five in physical size corresponds to three magnitudes and
such an improvement will, optimistically, be provided by the Large
Synoptic Survey Telescope (LSST) with first light in 2023. Since
events similar to 2008~TC$_3$ will still be rare and the timescales
short, it is of the utmost importance to identify virtually all impactors
and do that as early as possible, that is, immediately after their
discovery, to allow for spectroscopic follow-up. Another challenge
will be to obtain useful spectroscopic data of small, fast-moving
impactors; obtaining visible spectroscopy of 2008~TC$_3$ was a
borderline case with a four-meter-class telescope.

An alternative approach to determine the compositional structure of
the asteroid belt is to focus on meteorite falls and use their
predicted trajectories prior to the impact to identify the most likely
source region in the asteroid belt.  \citet{GranvikBrown} use the
observed trajectories of 25 meteorite-producing fireballs published to
date to associate meteorite types to their most likely escape routes
from the asteroid belt and the cometary region. We note that while
this approach is indirect, because it lacks the spectroscopic
observations prior to the impact, it results in substantially higher
statistics per unit time.

Long-term (timescales ranging from months to centuries) asteroid
impact monitoring is carried out by Jet Propulsion Laboratory's
(JPL's) Sentry\footnote{\url{http://neo.jpl.nasa.gov/risk}.} system
\citep{2001DPS....33.4108C} and the University of Pisa's close-approach monitoring system
CLOMON2\footnote{\url{http://newton.dm.unipi.it/neodys}.} \citep{1999DPS....31.2806C}. These automated systems are based on a
similar algorithm
\citep{2002aste.book...55M,2005Icar..173..362M,2015aste.book..815F}
and run in parallel. The results are continuously compared to identify
problematic cases that have to be scrutinized in greater detail by
human operators.

Whereas the long-term impact monitoring is well-established, dedicated
short-term (from hours to weeks) impact monitoring has only recently
started receiving serious attention. \citet{2015Icar..258...18F}
implemented an automated
system\footnote{\label{FCMJPL}\url{https://cneos.jpl.nasa.gov/scout/\#/}.}
\citep{2016DPS....4830503F} to regularly analyze the objects on the MPC's NEOCP with systematic
ranging. \citet{Spoto2018} have developed a short-arc orbit
computation method that also analyzes the objects on the NEOCP with
systematic ranging. Also \spotter{} relies on the observations on the
NEOCP, but the implementation and the orbit-computation method are
different. As in the case for long-term impact monitoring, comparing
the results of two or more systems doing the same analysis but using
different techniques provides assurance that the results are
valid. The very time-critical science described above underscores the
need for accurate and complete monitoring.
In addition to scientific utility, small impactors such as 2008~TC$_3$
and 2014~AA are valuable in comparing the predicted and actual impact
location and time. These small, non-hazardous impactors thus allow the
community to verify that the impact-monitoring systems work correctly.

\spotter{} also monitors for other types of interesting objects such
as asteroids temporarily captured by the Earth-Moon system.
\citet{2012Icar..218..262G} predict that there is a one-meter-diameter
or larger NEO temporarily orbiting the Earth at any given time. The
population of temporarily-captured natural Earth satellites (NES) are
challenging to discover with current surveys because they are small in
size and move fast when at a sufficiently small distance
\citep{2014Icar..241..280B}. Asteroid 2006~RH$_{120}$
\citep{2009A&A...495..967K} is to date the only NES confirmed not to
be a man-made object. It is a few meters in diameter and
therefore less challenging to discover compared to smaller objects.

In what follows we will first describe the real data available through
NEOCP as well as the simulated data used for testing \spotter{}. Then
we describe the \spotter{} system and the overall characteristics of
the numerical methods used (with detailed descriptions to be found in
the appendix). Finally, we present a selection of results and discuss
their implications, and end with some conclusions and future avenues
for improvement.

\section{Data}\label{data:data}

The MPC is a clearinghouse for asteroid and comet astrometry obtained
by observers worldwide. Based on an automatically computed probability
of a discovered moving object being potentially a new NEO, it is placed
on MPC's NEOCP for confirmation through follow-up
observations. Systems such as \protect \spotter{} and JPL's Scout
system\footref{FCMJPL} monitor the NEOCP for astrometry on new
discoveries.

Once the orbital solution for a new object has been sufficiently well
constrained to secure its orbital classification, the MPC
issues a Minor Planet Electronic Circular (MPEC) where the discovery
is published, and the object is removed from the NEOCP. Following the
release of an MPEC long-term (typically up to 100 years into the
future) impact-probability computations are undertaken by Sentry and
CLOMON2.

\subsection{Objects on the NEOCP}

As the primary test data set, we use 2339 observation sets of 695
objects on the NEOCP between August 19 and October 12, 2015 (eight weeks). We
use the term "observation set"\ to refer to astrometric
observations for a single object posted on the NEOCP within the last
half an hour, that is, \spotter{} checks the NEOCP for updates every half
an hour by default.

Of the 695 objects that appeared on the NEOCP, 260 objects eventually
received an MPEC. The remaining 435 objects were not NEOs or 
other objects on peculiar orbits, did not receive enough
follow-up observations to reveal their true nature, were linked to
previously discovered asteroids, or were found to be artificial
satellites or image artifacts. Zero percent (36\%) of the 260 objects that
received an MPEC (of the 435 objects that did not receive an MPEC)
have only one observation set. Zero percent (36\%) have two observation sets, that is, the object is updated once with more observations. Two percent (16\%)
have three observation sets and 51\% (37\%) four observation sets. A
large part of the observations in our test data set were obtained by
the Panoramic Survey Telescope and Rapid Response System (Pan-STARRS) 1 (F51) as well as the Catalina Sky Survey (703) and the
Mt. Lemmon Survey (G96). The first observation set for an object
contains observations from only one observatory. Seventy-seven percent of the 695
objects have at least two observation sets, in which case for 91\% of
the objects the first two observation sets come from different
observatories. Fifty-five percent of the 695 objects have at least three observation
sets, in which case for 91\% of the objects the first three
observation sets come from three observatories. An observation set
typically contains three or four observations. The time span of an
observation set for Pan-STARRS 1 is typically 30--60~min, and for the
Catalina Sky Survey and the Mt. Lemmon Survey 15--30~min. The time
between two observation sets ranges from about 30~min to two days. The
total time span when including up to four observation sets is typically
between two hours and three days.

\subsection{Simulated Earth impactors}\label{data:simul}

Apart from the two known Earth impactors mentioned in the introduction,
there are no real impactors for testing \spotter{}. Therefore we tested
\spotter{} with 50 simulated objects that were randomly picked from
the simulated sample generated by \citet{2009Icar..203..472V}. We
generated synthetic astrometry by using the location and the typical,
but simplified, cadence of Pan-STARRS 1. For 30 simulated impactors we
generated three astrometric observations with a 0.01~day interval
between the observations about five days prior to the impact with the
Earth. Then we generated three more astrometric observations for the
next night, that is, four days before the impact. For the remaining 20
simulated impactors we did the same but generated four astrometric
observations instead of three.

\subsection{Geocentric objects}\label{data:geocentric}

Similarly to Earth impactors, we have very few known NESs. Therefore we
test the system's capability to identify NESs with one asteroid
temporarily captured by the Earth, 2006~RH$_{120}$, and also the space
observatory Spektr-R.
2006~RH$_{120}$ was a natural temporarily-captured orbiter, which
orbited the Earth from July 2006 until July 2007. For 2006~RH$_{120}$
we used 17 observations between September 16 and 17 2006 spanning 24 hours
divided into four tracklets. The heliocentric elements at the end of
the epoch used are $a=1.04$ AU, $e=0.031$, and $i=1.43^{\circ}$ and
the geocentric elements are $a=0.013$ AU, $e=0.56,$ and
$i=64.2^{\circ}$.
Spektr-R (or RadioAstron) is a Russian Earth-orbiting space
observatory \citep{2013ARep...57..153K}. For Spektr-R we used eight
observations on April 4 2016 spanning four hours divided into three
tracklets. The geocentric elements are $a=180,000$ km, $e=0.93$, and
$i=38.8^{\circ}$ at the end of the epoch used.

\section{Methods}

\subsection{Statistical ranging}

In statistical orbital ranging
\citep{2001Icar..154..412V,2001CeMDA..81...93M}, the typically
non-Gaussian orbital-element probability density is examined using
Monte Carlo selection of orbits in orbital-element space using the
Bayesian formalism. In practice, the statistical ranging method
proceeds in the following way. Two observations are selected from the
data set and random deviates are added to the four plane-of-sky
coordinates to mimic observational noise. Next, topocentric distances
are generated for the two observation dates. The noisy sky-plane
coordinates and the two topocentric distances are transformed, by
accounting for the observatory's heliocentric coordinates at the
observation dates, into two heliocentric position vectors that are
mapped into the phase space of a Cartesian state vector, which is
equivalent to a set of orbital elements (e.g., Keplerian
elements). The proposed orbit is then used to compute ephemerides for
all observation dates. In Monte Carlo (MC) ranging the proposed orbit
is accepted if it produces acceptable sky-plane residuals and the
$\Delta\chi^2$ is small enough with respect to the until-then best-fit
orbit. The two other ranging variants using Markov chains require a
burn-in phase, which ends when, for random walk
\citep{2016P&SS..123...95M}, the first orbit with acceptable
$\Delta\chi^2$ is found, and for Markov-Chain Monte Carlo
\citep[MCMC,][]{2009M&PS...44.1897O}, the first orbit with acceptable
residuals at all dates is found. We use a constant prior
probability-density function (PDF) with Cartesian state vectors.

The different variants (MC, MCMC, and random-walk) of the statistical
orbital ranging method (Appendix \ref{appA}) used here have been
implemented in the open-source orbit-computation software package
OpenOrb. The ranging method in general has been
tested with several large-scale applications, demonstrating its wide
applicability to various observational data \citep[see][and references
  therein]{2015IAUGA..2256438V}.

The systematic ranging used by Scout \citep{2015Icar..258...18F} scans
a grid in the space of topocentric range $\rho$ and range rate
$\dot{\rho}$. For each grid point a fit in right ascension and
declination as well as their time derivatives ($\alpha$, $\delta$,
$\dot{\alpha}$, $\dot{\delta}$) is computed by minimizing a cost
function created from the observed-computed residuals and, combined
with the topocentric range and range rate, these six are equivalent to
a complete set of orbital elements. \citet{2015Icar..258...18F} also
tested different priors to be used with their analysis and concluded
that a uniform prior produces the best results. The choice of
an adequate prior distribution is critical to ensuring that impacting
asteroids are identified as early as possible. In what follows we will
use the uniform prior because that has been our default for many years
and now also \citet{2015Icar..258...18F} have shown that it is indeed a 
good choice. An alternative
implementation of the systematic ranging was recently proposed by
\citet{Spoto2018}.

\subsection{\protect \spotter{} system description}\label{method:descr}

The \protect \spotter{} system evaluates the Earth-impact probability
and various other characteristics of objects on the NEOCP. The
automated system works in general as follows. \protect \spotter{}
checks the NEOCP every 30~min for new objects and old objects with new
data. We note that the frequency of checks is a tunable parameter and
can be changed, but we have empirically found out that 30~min is a
reasonable update frequency for our current monitoring purposes.

For each observation set the different ranging methods are run (with a
timeout of twenty minutes) in the following order until the requested
number of sample orbits are computed: Monte Carlo (MC), Markov-Chain
Monte Carlo (MCMC), and random-walk ranging (Appendix
\ref{appA}). Previous orbit solutions are provided as input to ranging
to constrain the initial range bounds and speed up the
computations. For each observation set that results in the requested
number of accepted sample orbits, the orbital-element PDF is
propagated forward in time for seven days to derive the impact parameter
distribution. We note that also the length of the forward propagation
is a configuration parameter that can be changed if deemed
necessary. The computing time required increases roughly linearly with the length
of the forward propagation. The time needed to identify impactors can be reduced
by dividing the integrations according to a logarithmic sequence, say, one, three, ten, and 30 days,
so that, for example, the alert for an asteroid impacting within a day can be sent out after the first
integration rather than waiting for the 30~day integration to finish. The impact probability is computed as the weighted fraction
of impact orbits. Similarly, the probability of the object being
geocentric is computed by transforming the heliocentric orbital
elements to geocentric orbital elements, and computing the ratio of
those orbits that have geocentric $e < 1$.

The \protect \spotter{} system sends out a notification by e-mail of
all cases where the impact probability or the probability for an
object being captured in an orbit about the Earth is greater than
$10^{-4}$. We will next explain how this critical value and the
required number of sample orbits have been determined.

\section{Results}\label{results}

\subsection{Configuration parameters}\label{results:verif}

\begin{figure}[!ht]
\centering
\includegraphics[width=\columnwidth]{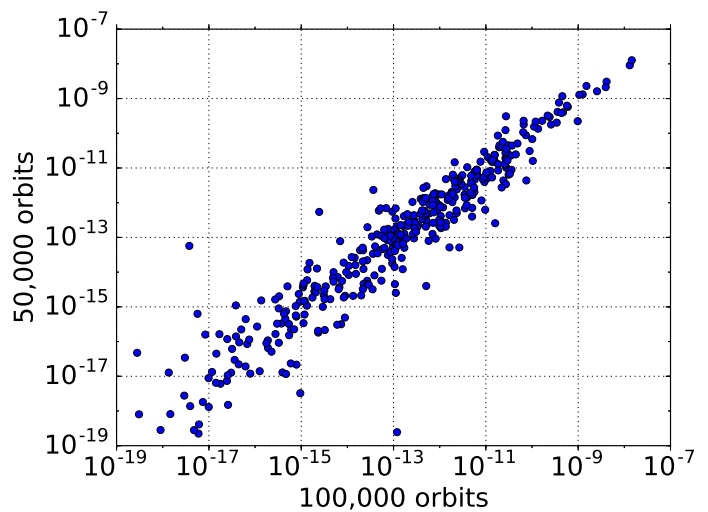}
\caption{Impact probabilities for 398 objects with 100,000 and 50,000
  orbits as the number of required sample orbits.}\label{orbitNprobs}
\end{figure}

We used conservative values for the many adjustable configuration
parameters in OpenOrb to be less sensitive to outliers in the first
data sets and to allow for the widest possible orbital-element
distribution. For the astrometric uncertainty we therefore assumed one arcsecond for all observatories. The configuration parameter that has the largest
influence on the computing time is the required number of sample
orbits. Since the number of sample orbits also directly affects the
reliability of the results, we decided to carry out tests to find a
suitable number. After 50,000 orbits the probabilities start to
converge when gradually increasing the number of required sample
orbits towards 100,000 (Fig.~\ref{orbitNprobs}). We therefore fixed the
number of required sample orbits to 50,000, because this is a good
compromise between computing time and the consistency of the results.

\subsection{Earth impactors}\label{res:simul}

\begin{figure}[!ht]
\centering
\includegraphics[width=\columnwidth]{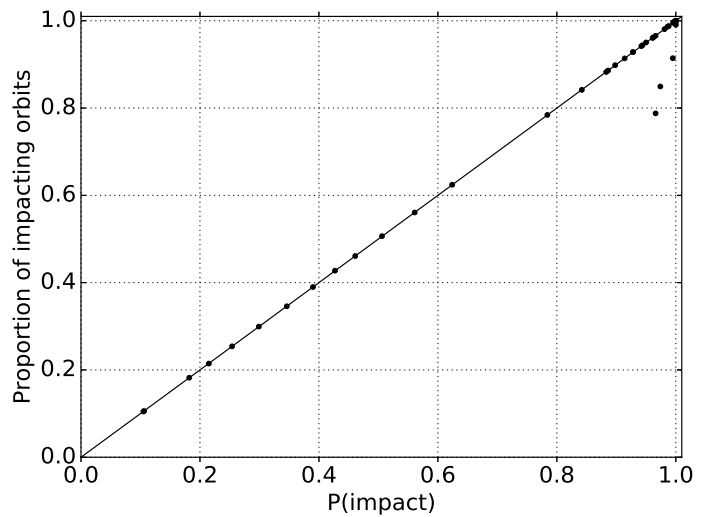}
\caption{For 50 simulated impactors, the impact probability versus the
  fraction of impacting orbits. The solid black line is included for
  reference and corresponds to a 100\% correlation between the
  fraction of impacting orbits and the impact
  probability.}\label{simulScatter}
\end{figure}

We tested the system with the two known Earth impactors (2008~TC$_3$ and
2014~AA) and 50 objects from the simulated sample generated by
\citet{2009Icar..203..472V}. For the latter we generated synthetic
astrometry for two consecutive nights as explained in
Sect.~\ref{data:simul}. To compute the sample orbits the MC ranging was
run twice: first with the observations of the first night, and then
with the observations of both nights initializing the analysis with
the orbits from the first ranging. The orbital-element probability
density function resulting from the ranging is propagated forward in
time as explained in Sect.~\ref{method:descr}. We found a strong
correlation between the impact probability and the number of impacting
orbits for the simulated impactors (Fig.~\ref{simulScatter}). This
implies that high impact probabilities are based on a statistically
meaningful number of sample orbits, and are thus unlikely to be false
alarms. We note that we do not here 
account for the fact that the typical false alarm is caused by 
a so-called outlier observation, an astrometric observation with the corresponding residual substantially larger than the expected astrometric uncertainty. The smallest impact probability is 10\%, ten objects have a
probability less than 50\%, and 32 objects have a probability greater
than 90\%. For the nights with four astrometric observations the
impact probabilities do not increase systematically compared to the
nights with three observations.

\begin{figure}[!ht]
\centering
\includegraphics[width=\columnwidth]{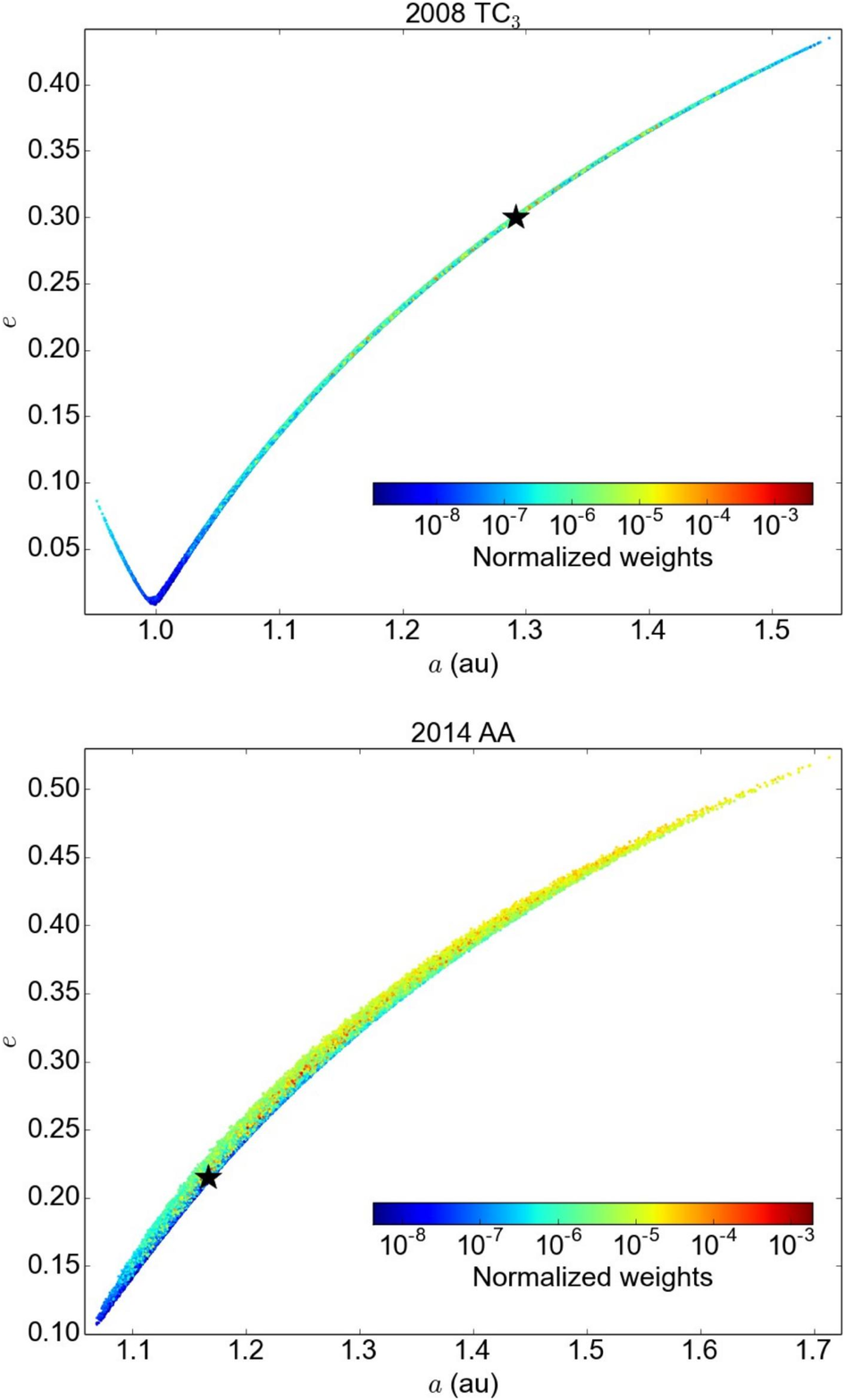}
\caption[]{Two-dimensional marginal probability densities for semi-major axis and
  eccentricity for the impactors 2008~TC$_3$ (epoch 54745.8 MJD) and 2014~AA (epoch 56658.3 MJD) based on the 
  first seven observations for both objects. The "true" orbital elements for the chosen epoch
  based on all available data as reported by JPL's HORIZONS system\footnotemark\ are marked with a
  star. We note that the estimate for 2014~AA also includes infrasound data \citep{2016Icar..274..327F}.}\label{impactors2D}
\end{figure}

When testing \spotter{} with the two known impactors, 2008~TC$_3$ and
2014~AA, we use two tracklets built from the first data sets
available. For 2008~TC$_3$ we use the first observations obtained by
R.~Kowalski: for the first tracklet four observations and for the
second tracklet three additional observations.  For 2014~AA we use
three observations for the first tracklet and the remaining four
observations available for the second tracklet.  For the first
tracklet the ranging takes one minute and for the second tracklet
ten minutes for both objects. With the addition of the second
tracklet the propagation takes a considerably longer time, an hour
for 2008~TC$_3$ and an hour and a half for 2014~AA), because due to many
impacting orbits the integration steps become smaller.  For both
objects the impact probabilities are negligible (less than one in ten billion) when using only the first
tracklet with a uniform prior, but increase to 81\% and 60\%, respectively, with Jeffreys' prior. The apparent discrepancy with the results obtained with a uniform prior by \citet{2015Icar..258...18F} will be studied in detail in a forthcoming paper. With the addition of the second tracklet the impact probabilities
increase to
92\% for 2008~TC$_3$ and to 80\% for 2014~AA. The two-dimensional (2D) marginal
probability densities for semi-major axis and eccentricity for
2008~TC$_3$ and 2014~AA are shown in Fig.~\ref{impactors2D}. The
PDFs resulting from the ranging method correctly cover the region
where the "true" orbit based on all available data (marked with a
star) resides.

\begin{figure}[!ht]
\centering
\includegraphics[width=\columnwidth]{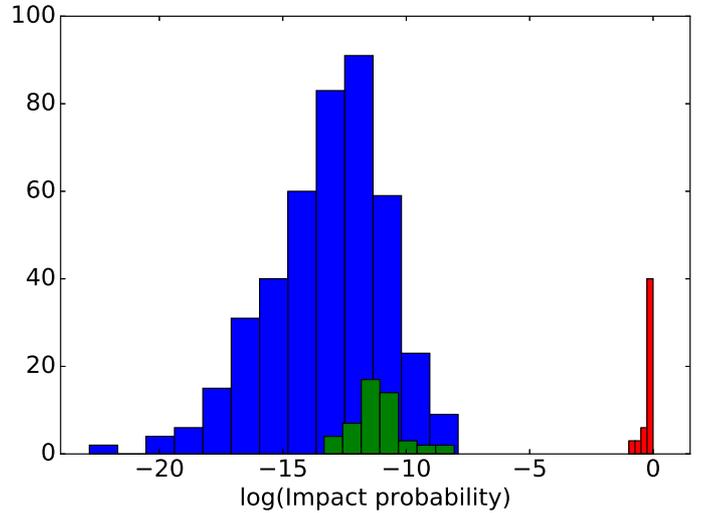}
\caption{Impact probabilities for NEOCP objects during eight weeks of
  continuous monitoring (blue), for 50 simulated impactors using one
  observation set (green), and for 2008~TC$_3$, 2014~AA, and 50
  simulated impactors using two observation sets from consecutive
  nights (red).}\label{probHisto}
\end{figure}

In Fig.~\ref{probHisto} the red histogram on the right represents
2008~TC$_3$, 2014~AA, and 50 simulated objects using two observation
sets from consecutive nights. The green histogram represents 50
simulated objects using only one observation set. Hence it is clear
that the substantially increased impact probability when adding a 
single data set to the discovery data set is typical for all 
impactors rather than being a random occurence for 2008~TC$_3$ and 
2014~AA.
\footnotetext{\url{https://ssd.jpl.nasa.gov/horizons.cgi}.}

\subsection{Geocentric objects}\label{res:geocentric}

In addition to potential Earth-impacting objects our system also
monitors for Earth's natural temporarily-captured satellites, that is,
objects on elliptic, geocentric orbits (hereafter just geocentric
orbits). Here we present results of tests with 2006~RH$_{120}$, an
asteroid temporarily captured by the Earth, and the space observatory
Spektr-R.

For 2006 RH$_{120}$ the first tracklet (four observations during three
minutes) produces only a negligible probability for the object being
geocentric, the second tracklet (nine observations during seven hours)
a 23\% probability, the third tracklet (14 observations during seven
hours) a 97\% probability, and the fourth tracklet (17 observations
during 24 hours) a 100\% probability. For Spektr-R the first tracklet
(four observations during one hour) gives a 24\% probability, the
second tracklet (six observations during three hours) a 97\%
probability, and the third tracklet (eight observations during four
hours) a 100\% probability.
From these two examples we see that our system recognizes a
geocentric object only if both the time span and number of observations
are sufficient.

\subsection{NEOCP objects}\label{results:test}

Having verified that \spotter{} can identify Earth-impacting asteroids
as well as objects on elliptical, geocentric orbits, we now present
results of eight weeks of continuous operations between August 18 and October
12 2015 for objects on the NEOCP. For each object \spotter{}
calculates the probability that the object will impact the Earth
within a week as well as the probability that the object is on an
elliptic, geocentric orbit. In addition, we compute an estimate for
the perihelion distance and NEO class. We compare these results with
those given by MPC.

We use 2339 observation sets of 695 objects on the NEOCP.  To derive
the impact parameter distribution the ranging is run for each
observation set and the resulting orbital element probability density
function is propagated forward in time for seven days as explained in
Sect.~\ref{method:descr}.  The ranging for the first observation set
takes on average one minute with root-mean-square (RMS) deviation of three minutes, for the
second observation set three minutes (RMS five minutes), for the third
observation set six minutes (RMS seven minutes), and for the fourth
observation set seven minutes (RMS eight minutes). The ranging gradually slows
down or fails for observation sets with larger time spans and number of
observations. If the ranging fails it is mostly because the timeout of
20 minutes is exceeded or because no acceptable sample orbits are
found (e.g., because the residuals are too large compared to the
assumed astrometric uncertainty). The time required for the
propagation stays typically around 1.5 minutes.

A non-zero impact probability is computed for 424 observation sets
(and in accordance with \citet{2006Icar..184..289V} almost always for
the first observation set of the object). The blue histogram in
Fig.~\ref{probHisto} shows the distribution of all non-zero impact
probabilities computed by \spotter{} during the eight weeks of
continuous monitoring. The ten highest probabilities in the blue
histogram fall between $7.7\times10^{-10}$ and
$2.3\times10^{-8}$. There is a clear gap between, on one hand, the
group formed by the 50 simulated impactors and the two real impactors
and, on the other hand, the group formed by the NEOCP objects and the
simulated impactors with only one observation set. It thus seems that
impactors are clearly identified already based on only two observation
sets, and \spotter{} produces very few, if any, false alarms.

A non-zero probability for the object being on an elliptic, geocentric
orbit is computed for 727 observation sets.  Some fraction of these
objects correspond to artificial Earth-orbiting satellites. In
particular, \spotter{} has computed a non-zero probability for a
geocentric orbit for all artificial objects that have appeared on
NEOCP and that have been identified as such on the various discussion
forums.

\begin{figure}[!ht]
\centering
\includegraphics[width=\columnwidth]{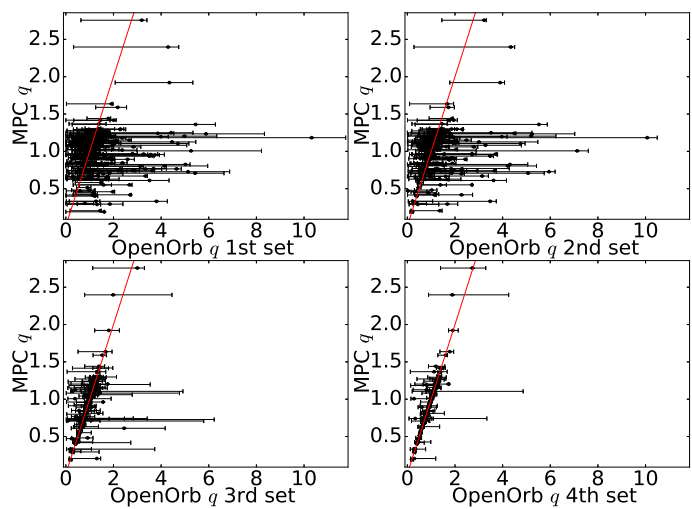}
\caption{Perihelion distance as estimated by \spotter{} and by MPC. The
  MPC estimate is based on all available data whereas the \spotter{}
  estimate only includes parts of it as described below each
  plot. The solid red line is included for
  reference and corresponds to a 100\% correlation between the
  perihelion distance as estimated by \spotter{} and by MPC.}\label{qFig}
\end{figure}

\begin{figure}[!ht]
  \centering
  \includegraphics[width=\columnwidth]{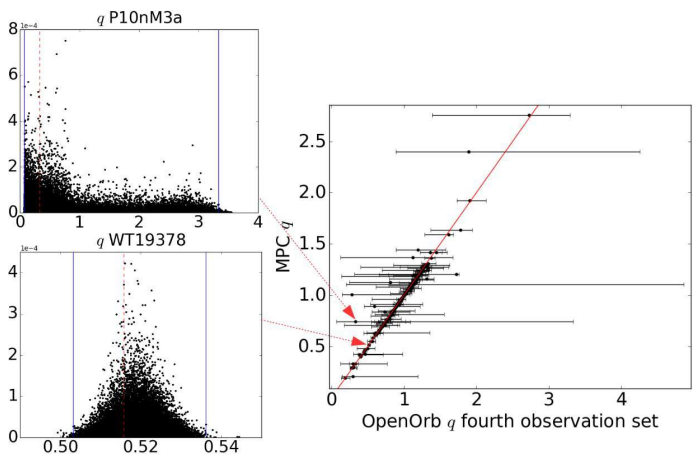}
  \caption{Two example distributions of the perihelion
    distance. The red vertical dashed line corresponds to the maximum likelihood and the blue
    vertical lines define the 1$\sigma$ confidence interval.}\label{4sets}
\end{figure}

The evolution of the perihelion-distance estimate with the increasing
number of observation sets for the 260 objects that received an MPEC
is shown in Fig.~\ref{qFig}. The error bars define the 1$\sigma$
confidence interval. The accuracy increases with each new observation
set as expected. Figure~\ref{4sets} shows examples of the
perihelion-distance distributions for two objects. We note the
substantially non-Gaussian distribution in the top left plot compared to the Gaussian distribution in the bottom left plot.

\begin{figure}[!ht]
\centering
\includegraphics[width=\columnwidth]{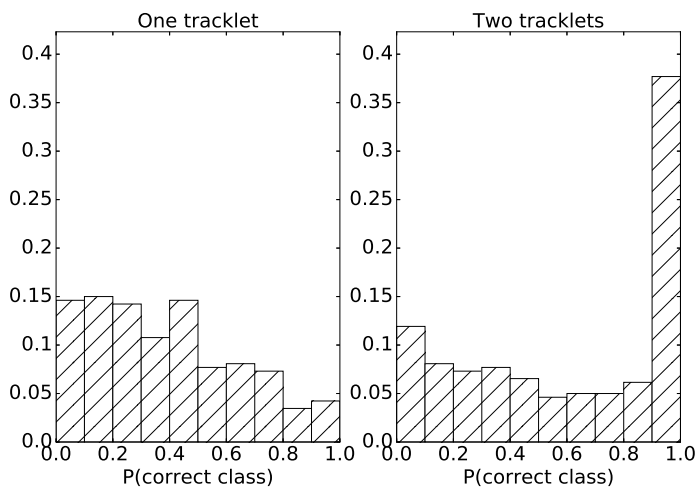}
\caption{Fractional distribution of the probabilities (as
  estimated by \spotter{}) to belong to the class reported by the MPC
  when using one or two observation sets.}\label{NEOclass}
\end{figure}

\spotter{} predicts the NEO class (Amor, Apollo, Aten,
Aethra/Mars-crosser) correctly (i.e., \spotter{} estimates the class
to be the one given by MPC with over 50\% confidence) for 40\% of the
260 objects in our test data set when using only the first observation
set and 63\% when using two observation sets (Fig.~\ref{NEOclass}).
In the rightmost bar in the plot for two tracklets we see that the
correct class is predicted for $0.377*260=98$ objects with at least a
90\% probability.

\begin{figure}[!ht]     
\centering
\includegraphics[width=\columnwidth]{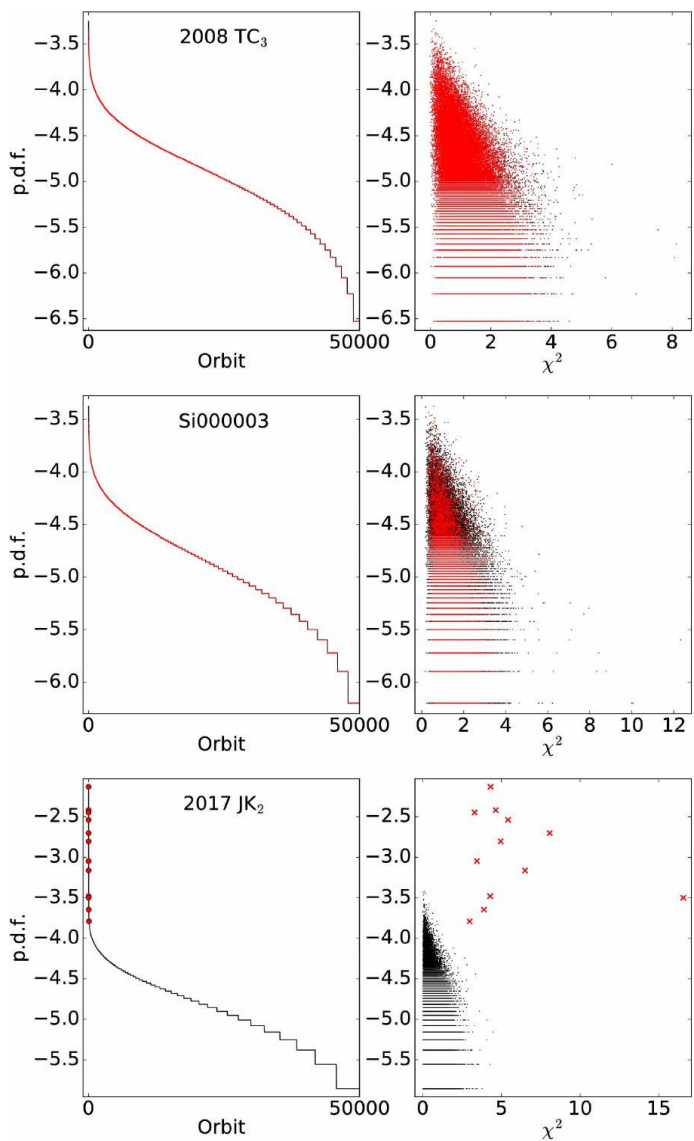}
\caption{Distribution of PDFs and $\chi^2$ values for 2008~TC$_3$,
  Si000003, and 2017~JK$_2$ resulting from one MCMC ranging run. The
  red symbols represent the impacting orbits.}\label{chi2}
\end{figure}

In rare cases (one object out of a thousand based on 13 months of
operations) \protect \spotter{} erroneously computes a high impact
probability based on a small number of impacting orbits, as in the
MCMC ranging run for 2017~JK$_2$ (Fig.~\ref{chi2}). In that case, only
12 impacting orbits produce an impact probability of 2\% although all
the impacting orbits have large $\chi^2$ values. These impacting
orbits also correspond to very small topocentric distances at the
orbit computation epoch, which is typically the mid-point of the
observational time span. These impacting orbits thus comprise the
low-likelihood tail of the orbit distribution. In MCMC ranging the
weight of a sample orbit is computed based on the repetitions, that
is, the inability of the proposed orbit to be accepted when starting
from the sample orbit in question. A viable scenario therefore is
that, in some rare cases, the sample orbits in the tail get an
unrealistically large number of repetitions, because there is
essentially just one narrow path towards higher-likelihood orbits and
the algorithm requires an unusually long time to find it. Since a
large fraction of these close-in orbit solutions typically impact the
Earth, the impact probability is also substantial despite the fact
that the solutions have low likelihoods based on their $\chi^2$
values. The other two examples in Fig.~\ref{chi2} are the real
impactor 2008~TC$_3$ and the simulated impactor Si000003 from
\citet{2009Icar..203..472V}. For 2008~TC$_3$ we get an impact
probability of 98\% with 48,293 impacting orbits (out of 50,000
orbits), and for Si000003 an impact probability of 24\% with 11,691
impacting orbits.

\section{Conclusion}\label{sec:conclusion}

While \spotter{} correctly predicts the impacts of the two asteroids
discovered before impacting the Earth, 2008~TC$_3$ and 2014~AA, the
impact probabilities for objects on the NEOCP computed by \spotter{}
are always less than one in ten million, as it should be for non-impactors
typically found on NEOCP. In addition, \spotter{} rarely produces
false alarms of impacting asteroids. As we gain experience from real
objects we will be able to apply a threshold for an impact probability
that should trigger follow-up efforts or, at least, close scrutiny of
the data.

We note that, for the two known impactors (2008~TC$_3$ and 2014~AA)
and 50 simulated impactors, a significant impact probability is
achieved only using two tracklets while using only one tracklet
results in a negligible impact probability. As a consequence our
system might not be able to identify an impactor based on the very
first data set available. It may be possible to increase the impact
probability computed for the first observation set by using a uniform
prior in polar coordinates \citep{2015Icar..258...18F}. We
leave the implementation and testing of the alternative prior(s) to
future works that should also provide a quantitative comparison of the
existing short-term impact-monitoring systems. Such a comparison would
provide end-users with a fact-based list of pros and cons of the
different systems.

\begin{acknowledgements}
We thank Federica Spoto (reviewer) and Davide Farnocchia for comments and suggestions that helped improve the paper.
OS acknowledges funding from the Waldemar von \mbox{Frenckell} foundation. 
MG acknowledges funding from the Academy of Finland (grant \#299543).
\end{acknowledgements}

\bibliographystyle{aa} 
\bibliography{AA_2018_32747}

\begin{appendix}

\section{Statistical orbital ranging}\label{appA}

\subsection{Inverse problem of orbit computation}

Within the Bayesian framework, the orbital-element probability density
function (PDF)
is\begin{equation}\label{A1}
p_\mathrm{p}(\boldsymbol{P})\propto p_\mathrm{pr}(\boldsymbol{P})p_\mathrm{\epsilon}(\Delta\boldsymbol{\mathit{\Psi}}(\boldsymbol{P})),
\end{equation}
where the prior PDF $p_\mathrm{pr}$ is constant and the
observational error PDF $p_\mathrm{\epsilon}$ is evaluated for the
observed–computed sky-plane residuals
$\Delta\boldsymbol{\mathit{\Psi}}(\boldsymbol{P})$ and is typically
assumed to be Gaussian \citep{1993Icar..104..255M}. The parameters $\boldsymbol{P}$
describe the orbital elements of an asteroid at a given epoch
$t_0$. For Keplerian orbital elements $\boldsymbol{P} = (a, e, i,
\Omega, \omega, M_0)^T$. For Cartesian elements, $\boldsymbol{P} = (X,
Y, Z, \dot{X}, \dot{Y}, \dot{Z})^T$ , where, in a given reference
frame, the vectors $(X, Y, Z)^T$ and $(\dot{X}, \dot{Y}, \dot{Z})^T$
denote the position and velocity, respectively.

\subsection{Statistical ranging}\label{originalR}

In statistical orbital ranging
\citep{2001Icar..154..412V,2001CeMDA..81...93M}, the orbital-element
PDF is examined using Monte Carlo selection of orbits
in orbital-element space in the following way:
\begin{itemize}
\item Select two observations (usually the first and the last) from
  the data set.
\item To vary the topocentric coordinates R.A. and Dec., and the
  topocentric range of the two observations introduce six uniform
  random deviates mimicking uncertainties in the coordinates.
\item Compute new Cartesian positions at the dates.
\item For the proposed orbit, compute candidate orbital elements,
  R.A. and Dec. for all observation dates, and $\Delta\chi^2$.
\item 
Let
\begin{eqnarray}\label{rw1}
& & \Delta \chi^2(\boldsymbol{P})=\chi^2(\boldsymbol{P})-\chi^2(\boldsymbol{P}_0) \,\mathrm{, and} \\
& & \chi^2(\boldsymbol{P})=\Delta\boldsymbol{\mathit{\Psi}}^T(\boldsymbol{P})\Lambda^{-1}\Delta\boldsymbol{\mathit{\Psi}}(\boldsymbol{P}) \nonumber\,,
\end{eqnarray}
where $\Lambda$ is the covariance matrix for the observational errors and $\boldsymbol{P}_0$ specifies a reference orbital
solution. Notice that, for linear models and Gaussian PDFs, the
definition of Eq. \ref{rw1} yields the well-known result
\begin{equation}\label{rw2}
\Delta \chi^2(\boldsymbol{P})=(\boldsymbol{P}-\boldsymbol{P}_0)^T\Sigma^{-1}(\boldsymbol{P}_0)(\boldsymbol{P}-\boldsymbol{P}_0),
\end{equation}
where $\boldsymbol{P}_0$ denotes the linear least-squares orbital
solution.

In MC ranging proposed orbits are accepted if they produce acceptable
sky-plane residuals: the $\Delta\chi^2$ value of the residuals is
below a given threshold (we use 30), and if the residuals at all observation dates are smaller than given cutoff values (for
$\Delta\alpha_{max}$ and $\Delta\delta_{max}$ we use 4 arcsecs). For
the acceptance criteria of the other ranging variants see
Sects. \ref{MCMCR} and \ref{RWR}.

\item If accepted, assign a statistical weight based on $\chi^2(\boldsymbol{P})$ describing the orbital-element PDF of the
  sample orbit:
  \begin{equation}
p_\mathrm{p}(\boldsymbol{P})\propto\mathrm{exp}\left[-\dfrac{1}{2}\chi^2(\boldsymbol{P})\right]\,.
\end{equation}
  
\item The procedure is repeated up to $10^7$ times resulting in up to 50,000
  accepted sample orbits. Both of these are adjustable
  parameters and their current values have been found empirically (see
  Sect.~\ref{results:verif}).
\end{itemize}

\subsection{MCMC ranging}\label{MCMCR}

Markov-chain Monte Carlo (MCMC) ranging \citep{2009M&PS...44.1897O} is
initiated with the selection of two observations from the full set of
observations: typically, the first and the last observation are
selected, denoted by A and B. Orbital-element sampling is then carried
out with the help of the corresponding topocentric ranges ($\rho_A$,
$\rho_B$), R.A.s ($\alpha_A$, $\alpha_B$), and Dec.s ($\delta_A$,
$\delta_B$). These two spherical positions, by accounting for the
light time, give the Cartesian positions of the object at two
ephemeris dates. The two Cartesian positions correspond to a single,
unambiguous orbit passing through the positions at the given dates.

For sampling the orbital-element PDF we utilize the Metropolis-Hastings (MH)
algorithm. The MH algorithm is based on the computation of the ratio
$a_r$:
\begin{equation}\label{MHar}
a_r=\frac{p_\mathrm{p}(\boldsymbol{P'})p_\mathrm{t}(\boldsymbol{P}_j;\boldsymbol{P'})}{p_\mathrm{p}(\boldsymbol{P}_j)p_\mathrm{t}(\boldsymbol{P'};\boldsymbol{P}_j)}=\frac{p_\mathrm{p}(\boldsymbol{P'})p_\mathrm{t}(\boldsymbol{Q}_j;\boldsymbol{Q'})J_j}{p_\mathrm{p}(\boldsymbol{P}_j)p_\mathrm{t}(\boldsymbol{Q'};\boldsymbol{Q}_j)J'}=\frac{p_\mathrm{p}(\boldsymbol{P'})J_j}{p_\mathrm{p}(\boldsymbol{P}_j)J'},
\end{equation}
where $\boldsymbol{P}_j$ and $\boldsymbol{Q}_j$ denote the current orbital
elements and spherical positions, respectively, in a Markov chain, and
the primed symbols, $\boldsymbol{P'}$ and $\boldsymbol{Q'}$, denote
their proposals.  The proposal PDFs from $\boldsymbol{P}_j$ to
$\boldsymbol{P'}$ and from $\boldsymbol{Q}_j$ to $\boldsymbol{Q'}$ (t
stands for transition) are, respectively,
$p_\mathrm{t}(\boldsymbol{P'};\boldsymbol{P}_j)$ and
$p_\mathrm{t}(\boldsymbol{Q'};\boldsymbol{Q}_j)$.

The proposal $\boldsymbol{P'}$ is transformed to the space of two topocentric
spherical positions
$\boldsymbol{Q'}=(\rho_\mathrm{A}',\alpha_\mathrm{A}',\delta_\mathrm{A}',\rho_\mathrm{B}',\alpha_\mathrm{B}',\delta_\mathrm{B}')^\mathrm{T}$
resulting in a multivariate Gaussian proposal
PDF $p_\mathrm{t}(\boldsymbol{Q'};\boldsymbol{Q}_j)$. This
transformation introduces Jacobians $J_j$ and $J'$:
\begin{equation}\label{jacob}
J_j=\left | \frac{\partial \boldsymbol{Q}_j}{\partial \boldsymbol{P}_j} \right |,\quad J'=\left | \frac{\partial \boldsymbol{Q'}}{\partial \boldsymbol{P'}} \right |.
\end{equation}

The ratio $a_r$ in Eq. \ref{MHar} simplifies into its final form
because the proposal PDFs
$p_\mathrm{t}(\boldsymbol{Q_j};\boldsymbol{Q}')$) and
$p_\mathrm{t}(\boldsymbol{Q'};\boldsymbol{Q}_j)$) are symmetric.

The proposed elements $\boldsymbol{P'}$ are accepted with the
probability of min(1,$a_r$):
\begin{align}
\mathrm{If\ } a_r\geq 1; \mathrm{\ then\ } \boldsymbol{P}_{j+1}& = \boldsymbol{P'} \nonumber \\
\mathrm{If\ } a_r<1; \mathrm{\ then\ } \boldsymbol{P}_{j+1}& = \boldsymbol{P'} \mathrm{\ with \ probability\ } a_r \nonumber \\
\boldsymbol{P}_{j+1}& = \boldsymbol{P'} \mathrm{\ with \ probability\ } 1-a_r. \numberthis \label{acceptElem}
\end{align}

After a number of transitions in the so-called burn-in phase, the
Markov chain, in the case of success, converges to sample the target
PDF $p_\mathrm{p}$ when the first orbit with acceptable residuals
at all dates is found.

The posterior distribution is proportional to the number of repetitions 
of a given orbit $r(\boldsymbol{P})$ rather than on the $\chi^2(\boldsymbol{P})$ values:
\begin{equation}
p_\mathrm{p}(\boldsymbol{P}) \propto r(\boldsymbol{P})\,.
\end{equation}
That is, the difficulty to find an 
acceptable $\boldsymbol{P'}$ increases the probability of the current orbit 
$\boldsymbol{P}$.

\subsection{Random walk ranging}\label{RWR}

Instead of MCMC ranging, it can be advantageous to sample in the
entire phase-space regime below a given $\chi^2(\boldsymbol{P})$
level, assigning weights on the basis of the a posteriori probability
density value and the Jacobians presented above \citep[cf.,
  \citealt{2001Icar..154..412V},
  \citealt{2001CeMDA..81...93M}]{2016P&SS..123...95M}.

MCMC ranging can be modified for what we call random-walk ranging of
the phase space within a given $\Delta\chi^2$ level. First, we assign a
constant, nonzero PDF for the regime of acceptable orbital
elements and assign a zero or infinitesimal PDF value outside the
regime. MCMC sampling then returns a set of points that, after
convergence to sampling the phase space of acceptable orbital elements
(i.e., the burn-in phase ends when the first orbit with acceptable
$\Delta\chi^2$ is found), uniformly characterizes the acceptable
regime. Second, we assign the posterior PDF values $p_\mathrm{p}$
as the weights $w_j$ for the sample orbital elements
$\boldsymbol{P}_j$. In the case of ranging using the topocentric
spherical coordinates, the weights $w_j$ need to be further divided by
the proper Jacobian value $J_j$.

In random-walk ranging, uniformly sampling the phase space of the
acceptable orbital elements, the final weight factor for the sample
elements $\boldsymbol{P_j}$ in the Markov chain is
\begin{equation}\label{rw3}
w_j=\frac{r(\boldsymbol{P_j})}{J_j}\mathrm{exp}\left[-\dfrac{1}{2}\chi^2(\boldsymbol{P}_j)\right].
\end{equation}
We note that the same orbit can repeat itself in the chain, analogously
to MCMC ranging.

\end{appendix}

\end{document}